## NONLINEAR AND QUANTUM OPTICS

# Simulation of Interaction of Oriented *J* Aggregates with Resonance Laser Radiation

N. V. Vysotina[a, b], V. A. Malyshev[c, d], V. G. Maslov[a], L. A. Nesterov[a],
N. N. Rosanov[a, b], S. V. Fedorov[a, b], and A. N. Shatsev[a, b]

[a] *St. Petersburg State University of Information Technologies, Mechanics, and Optics, St. Petersburg, 197101 Russia*
[b] *Vavilov State Optical Institute, St. Petersburg, 199034 Russia*
[c] *Zernike Institute for Advanced Materials, University of Groningen, The Netherlands*
[d] *St. Petersburg State University, Peterhof, St. Petersburg, 198504 Russia*
*e-mail: nrosanov@yahoo.com*
Received October 29, 2009

**Abstract**—The interaction of laser radiation with single *J* aggregates of cyanine dyes is theoretically analyzed and numerically simulated. The quantum-mechanical calculations of the equilibrium geometry and the energies and intensities of the lowest singlet electronic transitions in pseudoisocyanine chloride and its linear (chain) oligomers are fulfilled. The data of these calculations can serve as parameters of the analyzed model of interaction of *J* aggregates with radiation in the one-particle density matrix approximation. This model takes into account relaxation processes, the annihilation of excitations at neighboring molecules, and inhomogeneous broadening. Assuming that the inhomogeneous broadening is absent, calculations demonstrate the existence of spatial bistability, molecular switching waves, and dissipative solitons. The effect of the inhomogeneous broadening and the radiation intensity on the effective coherence length in linear (chain) *J* aggregates is analyzed.

**DOI:** 10.1134/S0030400X10070180

## 1. INTRODUCTION

The resonance excitation of *J* aggregates of cyanine dyes by laser radiation occurs via the collective (excitonic) mechanism. This agrees with their intense fast (subpicosecond) optical response (both linear and nonlinear), which makes these low-dimensional structures promising for application in data processing [1] and laser engineering [2]. Of special interest is the possibility of applying them for molecular memory schemes. The bistability of a single *J* aggregate resonantly excited by radiation has been predicted and studied in theoretical works [3, 4]. Since the *J* aggregates of cyanine dyes consist of hundreds and thousands of molecules, it seems promising to create a miniature memory cell with a noticeably smaller size than that of the complete aggregate. Recent theoretical work [5] demonstrated that laser excitation of *J* aggregates can induce the formation of discrete dissipative solitons localized almost completely on one molecule. This is obviously the first example of molecular dissipative solitons and the first example of dissipative nanosolitons (about 1 nm in size). These circumstances actualize the problem of the more complete simulation of the interaction of oriented *J* aggregates of cyanine dyes with cw and pulsed laser radiation.

Full-scale simulation of the resonance interaction of multimolecular structures with laser radiation is difficult both due to the insufficient development of the theoretical model and due to calculation problems. The model proposed in [4] correlates three-level systems to individual aggregate molecules. The optical dynamics of molecules is described by the Bloch equations for one-molecule density matrices, which explicitly takes into account the intermolecular dipole–dipole interaction and the excitation annihilation. The model parameters are chosen based on the experimental data or evaluation considerations.

This model is restricted by the following factors. First, the method does not use modern methods of computational quantum chemistry, which allow one to sufficiently correctly determine the main parameters of individual dye molecules and their complexes, which are needed to set the model. Second, there are additional important factors, such as relaxation processes and inhomogeneous broadening (the spread in the molecular transition frequencies due to the interaction of molecules with the environment, which is observed even at low temperatures). The aim of this study is to develop a more complete model of the interaction of single *J* aggregates with laser radiation using a reasonable combination of first-principle (ab initio) or semiempirical (quantum-chemical) calculations and a simplified model based on one-particle density matrices, some parameters of which are derived from quantum-chemical calculations.

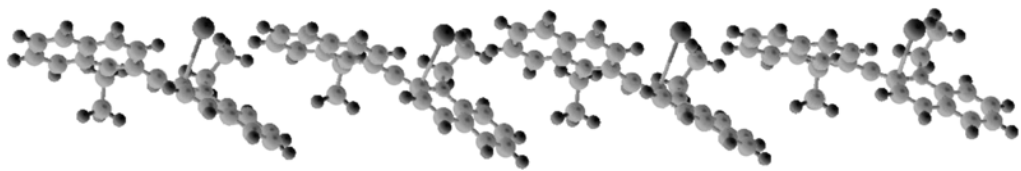

**Fig. 1.** Geometry of (PIC:Cl)$_4$ aggregate calculated by DFT/B3LYP/6-31G method. Hydrogen and chlorine atoms are darker than carbon and nitrogen atoms. Chlorine atoms are larger in size than hydrogen atoms.

## 2. QUANTUM-CHEMICAL CALCULATIONS OF ELECTRONIC TRANSITIONS

In this section, we present the results of calculations of the equilibrium geometry and the energies and intensities of the lowest singlet electronic transitions of pseudoisocyanine chloride (PIC:Cl) and its linear (chain) oligomers (PIC:Cl)$_N$. The equilibrium geometry was calculated for PIC:Cl, (PIC:Cl)$_2$, and (PIC:Cl)$_4$ by the DFT/B3LYP/6-31G density functional method using the GAMESS program [6]. The excitation energies and oscillator strengths of electronic transitions were calculated by the ZINDO/S-CI method using the program described in [7, 8]. It was assumed that the aggregate has an idealized geometry that differs from the calculated optimized geometry by the following assumptions: (i) all the dye molecules are identical to each other (we used the averaged geometry of molecules calculated by the DFT method for the (PIC:Cl)$_4$ aggregate); (ii) the aggregate is constructed by a parallel transfer of a PIC:Cl molecule onto the vector $m\mathbf{b}$, where the vector $\mathbf{b}$ was taken to be the same for all aggregates and was determined as an average shift between molecules, which was also calculated by the DFT method for the (PIC:Cl)$_4$ aggregate; and (iii) to avoid the calculation of unreal states of charge transfer from Cl$^-$ to PIC$^+$, we used F atoms instead of Cl in the same positions; i.e., the energies and oscillator strengths of transitions were actually calculated for the (PIC:F)$_N$ aggregates.

**Equilibrium geometry.** This geometry is shown in Fig. 1 by the example of (PIC:Cl)$_4$. The geometric parameters that characterize the calculated equilibrium geometry of the aggregate were found to be as follows: the displacement (norm of the vector $\mathbf{b}$) $a = 11.38$ Å and the angle between the normal to the molecule plane and the vector $\mathbf{b}$ of about 63°. The latter value cannot be determined exactly because the definition of the plane of the molecule in this case is not completely correct. Deviations of some atoms (even of atoms entering the π electronic system) from the planar structure are rather large. A stricter parameter is the angle between the transition dipole moment (see below) of the monomer molecule and the vector $\mathbf{b}$, which, in this case, is 27.8°.

**Energies and intensities of electronic transitions.** The excitation energies and oscillator strengths for a series of the lowest excited states of (PIC:F)$_N$ aggregates are listed in Table 1.

The effect of the summation of intensities in the long-wavelength component of the exciton-resonance multiplet is quite obvious. The oscillator strength for this transition approximately linearly depends on the number of molecules in the aggregate. It is important that the direction of the dipole moment for this transition is the same for all aggregates and coincides with that for the monomer molecules, i.e., forms an angle of 27.8° with the vector $\mathbf{b}$.

An analysis of the intensities of the other components of the long-wavelength exciton-resonance multiplet shows the existence of transitions with a very low ($f \sim 0.001$) and rather noticeable ($f > 0.1$) intensities. The latter have the same direction of the transition dipole moment as in the long-wavelength component (i.e., at an angle of 27.8° to the vector $\mathbf{b}$). Beginning from aggregates with $N = 4$, we observed a clear tendency for intense and weak transitions to alternate on the energy scale, which is even more pronounced in the case of aggregates with $N = 50$, the calculated data for which are shown in Table 2.

According to theoretical considerations (see, e.g., [9]), the oscillator strengths of exciton components of an ideal linear chain of identical molecules are described by the relation

$$f_k = f_0 \frac{1-(-1)^k}{N+1} \cot^2 \frac{\pi k}{2(N+1)}, \quad (1)$$

where $f_0$ is the oscillator strength of the corresponding transition of the monomer molecule, $N$ is the number of molecules in the chain, and $k$ is exciton component number. The values calculated by formula (1) using $f_0 = 1.37$ from Table 1 are presented in Table 2. A comparison of these values with the oscillator strengths calculated by exact quantum-chemical formulas shows good agreement.

## 3. INTERACTION OF *J* AGGREGATES WITH RADIATION

### 3.1. Equations for One-Particle Density Matrix

A *J* aggregate is simulated according to [4] by a linear chain of $N$ three-level molecules interacting with each other through an electromagnetic field. Two lower levels form an optical transition, which is quasi-resonance to the incident laser field. The third molecular level serves to describe the annihilation of two excitations at neighboring molecules: one of the excitations is deactivated and the other passes to the third level and then relaxes back to the second or to the first level [10]. As a result, either one or both excitations disappear from the system. The frequency of the transition from the ground (1) to the third (3) state is assumed to be approximately equal to the double frequency of the transition $1 \to 2$. The distance $a$ between molecules is much smaller than the radiation wavelength. The laser radiation frequency is $\omega_0$ and the radiation polarization is linear.

**Table 1.** Excitation energies (in cm$^{-1}$) and oscillator strengths $f$ of (PIC:F)$_N$ aggregates. The data are given only for the lowest excited singlet state and for the higher states composing the long-wavelength exciton-resonance multiplet

|          | $N=1$ | $N=2$ | $N=3$ | $N=4$ | $N=5$ | $N=6$ | $N=7$ | $N=8$ |
|----------|-------|-------|-------|-------|-------|-------|-------|-------|
| $S_1(f)$ | 20767 (1.3733) | 20950 (2.4080) | 21018 (3.6521) | 21145 (5.0232) | 21061 (6.1276) | 21047 (7.2459) | 21049 (8.5368) | 21028 (9.6244) |
| $S_2(f)$ |       | 21965 (0.3347) | 21695 (0.3705) | 21705 (0.0004) | 21547 (0.0008) | 21446 (0.0474) | 21360 (0.0161) | 21299 (0.0091) |
| $S_3(f)$ |       |       | 22215 (0.0406) | 22122 (0.2730) | 21955 (0.4640) | 21811 (0.5408) | 21700 (0.5029) | 21605 (0.6020) |
| $S_4(f)$ |       |       |       | 22333 (0.0009) | 22225 (0.0000) | 22102 (0.0030) | 21989 (0.0002) | 21886 (0.0001) |
| $S_5(f)$ |       |       |       |       | 22359 (0.0568) | 22294 (0.1024) | 22196 (0.1602) | 22111 (0.2054) |
| $S_6(f)$ |       |       |       |       |       | 22402 (0.0004) | 22337 (0.0025) | 22269 (0.0148) |
| $S_7(f)$ |       |       |       |       |       |       | 22429 (0.0160) | 22374 (0.0775) |
| $S_8(f)$ |       |       |       |       |       |       |       | 22439 (0.0033) |

We use an approach based on the one-particle density matrix [3, 4]. In the approximation of slowly varying amplitudes, the initial equations for the density matrix elements $\rho_{ik}$, $i, k = 1, 2, ..., N$ have the form

**Table 2.** Excitation energies and oscillator strengths of transitions to ten lowest excited singlet states of the (PIC:F)$_{50}$ aggregate

| No. | Excitation energy, eV | Oscillator strength $f$ | Oscillator strength $f$ calculated by formula (1) |
|-----|-----------------------|-------------------------|---------------------------------------------------|
| 1   | 2.7588 | 55.1045 | 56.60 |
| 2   | 2.7618 | 0.0010  | 0     |
| 3   | 2.7659 | 5.0993  | 6.26  |
| 4   | 2.7709 | 0.0000  | 0     |
| 5   | 2.7765 | 1.7007  | 2.23  |
| 6   | 2.7826 | 0.0000  | 0     |
| 7   | 2.7890 | 0.8164  | 1.12  |
| 8   | 2.7957 | 0.0000  | 0     |
| 9   | 2.8026 | 0.4656  | 0.66  |
| 10  | 2.8095 | 0.0000  | 0     |

$$\dot{\rho}_{11}^{(k)} = \frac{1}{4}\sum_{l=1}^{N}(\gamma_{lk} + i\Delta_{lk})R_l R_k^* - i\frac{\Omega}{4}R_k^* + \text{c.c.} \\ + \alpha_s \rho_{22}^{(k)}[\rho_{22}^{(k-1)} + \rho_{22}^{(k+1)}] + \Gamma_{31}\rho_{33}^{(k)} + \Gamma_{21}\rho_{22}^{(k)}, \quad (2a)$$

$$\dot{\rho}_{22}^{(k)} = -\frac{1}{4}\sum_{l=1}^{N}(\gamma_{lk} + i\Delta_{lk})R_l R_k^* + i\frac{\Omega}{4}R_k^* + \text{c.c.} \\ - 2\alpha_s \rho_{22}^{(k)}[\rho_{22}^{(k-1)} + \rho_{22}^{(k+1)}] + \Gamma_{32}\rho_{33}^{(k)} - \Gamma_{21}\rho_{22}^{(k)}, \quad (2b)$$

$$\dot{\rho}_{33}^{(k)} = \alpha_s \rho_{22}^{(k)}[\rho_{22}^{(k-1)} + \rho_{22}^{(k+1)}] - (\Gamma_{31} + \Gamma_{32})\rho_{33}^{(k)}, \quad (2c)$$

$$\dot{R}_k = -(\Gamma_\perp + i\Delta_k)R_k + \sum_{l=1}^{N}(\gamma_{lk} + i\Delta_{lk})R_l[\rho_{22}^{(k)} - \rho_{11}^{(k)}] \\ - i\Omega[\rho_{22}^{(k)} - \rho_{11}^{(k)}] - \alpha_s R_k[\rho_{22}^{(k-1)} + \rho_{22}^{(k+1)}], \quad (2d)$$

where the point in the left-hand sides of equations denotes the first derivative. The matrices $\gamma_{lk}$ and $\Delta_{lk}$ ($l \ne k$) describe the delayed dipole–dipole interaction of aggregate molecules,

$$\gamma_{lk} = \frac{\mu^2}{\hbar a^3} \left\{ \left[ k_0 a \frac{\cos(k_0 a |l-k|)}{|l-k|^2} - \frac{\sin(k_0 a |l-k|)}{|l-k|^3} \right] \right.$$
$$\times (1 - 3\cos^2\theta) + (k_0 a)^2 \frac{\sin(k_0 a |l-k|)}{|l-k|} \sin^2\theta \right\},$$
$$l \neq k,$$
$$\Delta_{lk} = \frac{\mu^2}{\hbar a^3} \left\{ \left[ \frac{\cos(k_0 a |l-k|)}{|l-k|^3} + k_0 a \frac{\sin(k_0 a |l-k|)}{|l-k|^2} \right] \right. \quad (3)$$
$$\times (1 - 3\cos^2\theta) - (k_0 a)^2 \frac{\cos(k_0 a |l-k|)}{|l-k|} \sin^2\theta \right\},$$
$$l \neq k, \quad \gamma_{ll} = 0, \quad \Delta_{ll} = 0.$$

The parameter $\Delta_k$ in (2d) denotes the resonance frequency detuning for the $k$th molecule,

$$\Delta_k = \omega_{21}^{(k)} - \omega_0 = \Delta_0 + \delta\Delta_k. \quad (4)$$

Here, $\Delta_0$ is the average detuning for the molecules in the chain and $\delta\Delta_k$ is its statistical spread, i.e., the normal (Gaussian) frequency distribution for transitions of individual molecules (independently of each other) with zero mean value and a specified dispersion (for each $k$th molecule, it happens independently of the values for the other molecules). The parameter $\Omega = \frac{\mu}{\hbar} E_i$ is the Rabi frequency proportional to the complex field amplitude, so that the real electric field strength in the incident wave is

$$\tilde{E} = \text{Re}[E_i(t)\exp(-i\omega_0 t)]. \quad (5)$$

The other parameters of the problem are the transition dipole moment matrix element $\mu$, the angle $\theta$ between the dipole moment and the chain axis, the relaxation constants $\gamma$ and $\Gamma$, and the exciton–exciton annihilation parameter $\alpha$. This approach is a priori valid if the excitation of molecules is weak as follows:

$$1 - \rho_{11}^{(k)} \ll 1 \quad (6)$$

(the diagonal density matrix elements are proportional to the populations of the corresponding levels).

To reduce the equation to a dimensionless form, it is convenient to use the parameter $\gamma_R$, which is natural for homogeneous distributions,

$$\gamma_R = 2\sum_{l-k=1}^{N/2} \gamma_{lk}. \quad (7)$$

We normalize the time, the resonance frequency detunings, the Rabi frequency, and the exciton annihilation parameter to the parameter $\gamma_R$,

$$t' = \gamma_R t, \quad \Delta_0' = \Delta_0/\gamma_R, \quad \delta\Delta_k' = \delta\Delta_k/\gamma_R,$$
$$\Omega' = \Omega/\gamma_R, \quad \alpha_s' = \alpha_s/\gamma_R. \quad (8)$$

The relaxation rates G are normalized in the same way. Below, primes are omitted.

The main parameters are given in [3, 4] on the assumption $\Gamma = 0$ in the absence of statistical spread of detunings. At the same time, to search for the conditions of existence of various localized structures in a wider class of molecular aggregates, it is also useful to analyze other values of the parameters. For our calculations, we chose the following parameters: the number of molecules in the aggregate $N = 80$–$400$, $k_0 a = 0.1$, $1 - 3\cos^2\theta = -1$, and $\Omega = 0.1$–$2$. The frequency detuning was chosen from the condition $\Delta_0 - \Delta_L = -10$, where $\Delta_L = 2\sum_{l-k=1}^{N/2} \Delta_{lk}/\gamma_R$ (for *J* aggregates, $\Delta_L < 0$, $|\Delta_L| \gg 1$). The characteristic spread of frequency detunings ranges from 0 (homogeneous broadening) to $0.2|\Delta_L|$; the relaxation parameters are $\Gamma_{21} = 1$, $\Gamma_\perp = 1.1$, $\Gamma_{31} = 0.01$, $\Gamma_{32} = 0.99$; and the exciton annihilation parameter is $\alpha = 41.1$.

In linear aggregates, the interaction of molecules at the chain edges with radiation considerably differs from the interaction of molecules in the middle of the chain mainly due to the two-exciton annihilation (the parameter $\alpha$). According to our calculations, the edge effects involve several tens of molecules; therefore, these effects can be neglected for the middle part of a sufficiently long chain. To eliminate them from calculations, we can assume that the chain is closed in a ring (periodic boundary conditions). However, although molecular aggregates with a ring geometry are known, their molecules differently interact with linearly polarized radiation (the Rabi frequency is proportional to the cosine of the angle between the directions of the molecular dipole moment and the radiation polarization). Therefore, to create identical conditions for different molecules, it is necessary to use radiation with, e.g., circular polarization. Below, in addition to a linear chain, we will consider ring aggregates because the calculations in this case are much simpler. In subsections 3.2–3.4, we analyze a case with a homogeneous broadening (when the frequency spread dispersion of the main molecular transition is $\sigma = 0$), while the effect of the frequency dispersion ($\sigma \neq 0$) is studied in subsection 3.5.

### 3.2. Bistability in Case of Homogeneous Broadening

Although bistability in the case of resonance interaction of molecular *J* aggregates with laser radiation is studied in sufficient detail [3, 4], we present these calculations in this paper because, first, model (2) contains additional relaxation parameters and, second, the results of these calculations are important for considerations below since the bistability is favorable for the existence of dissipative solitons [11].

**Ring-shaped *J* aggregate.** In this configuration, there exist uniform (over aggregate) states with identi-cal level populations and dipole moments (off-diagonal density matrix elements) for all the molecules. Bistability means that, depending on the initial conditions, there are stable uniform distributions with two different values of these parameters. For example, for a ring with $N = 120$ molecules at $\Omega \geq 1.045$, there is only one stable state, in which the relative population of the lower level is comparatively small ($\rho_{11} \approx 0.91$) and the relative population of the second level is comparatively high ($\rho_{22} \approx 0.0274$). In contrast, at $\Omega \leq 0.9035$, there is only one stable state with a high relative population of the lower level ($\rho_{11} \approx 0.99$) and a low population of the second level ($\rho_{22} \approx 0.005$). In the intermediate region $0.9035 \leq \Omega \leq 1.045$, these two states can be implemented depending on the initial conditions (Fig. 2). Note that the bistability region boundaries can strongly depend on the number of molecules *N* and are not universal in this respect.

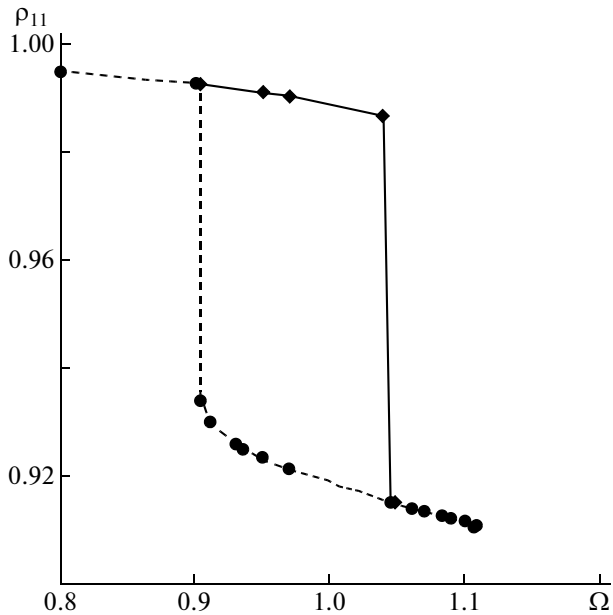

**Fig. 2.** Bistability of lower level populations for a ring-shaped *J* aggregate. $N = 120$.

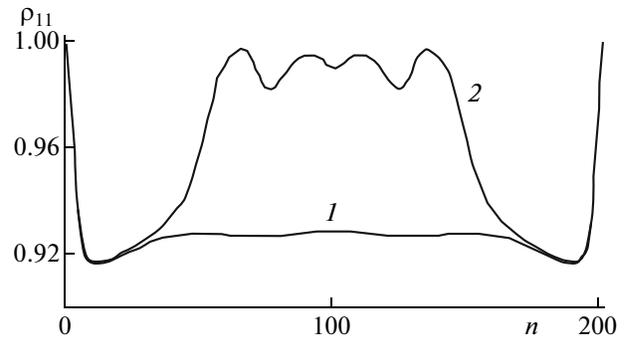

**Fig. 3.** Spatial bistability in a linear *J* aggregate. The upper curve shows the stable population of the lower level in the case of the initial population $\rho_{11} = 0.99$, and the lower curve is plotted for the initial population $\rho_{11} = 0.9$. The Rabi frequency is $\Omega = 0.94$; $N = 200$.

**Linear chain.** For finite chains, a bistability region also exists but with some limitations. As was mentioned above, a uniform distribution of level populations $\rho_{11}$ and $\rho_{22}$ in this case is impossible due to the edge effect. However, a spatial bistability can exist; i.e., two different profiles of the density matrix elements can be settled depending on the initial conditions. An approximately uniform distribution can exist in the middle of sufficiently long chains. An example of the spatial bistability is given in Fig. 3, which shows two stable profiles of the lower-level populations. It is seen that the edge effect manifests itself for about 50 molecules from each edge. One of the profiles corresponds to a comparatively low population of the lower level over almost the entire chain (except for its edges), while the second profile demonstrates a higher population of this level in the central part of the chain. This regime will be compared with the case of solitons in Subsection 3.4.

The spatial bistability, similar to the ordinary bistability, is implemented in a limited region of parameters. This region is shown in Fig. 4, which presents the dependence of the maximum (over the chain) population of the lower level on the Rabi frequency. In comparison with the ring-shaped aggregate, the stability region for a chain is shifted and narrowed. The occurrence of a given branch is determined by the initial conditions ($\rho_{11} = 0.99$ and $\rho_{22} = 0.004$ for one branch and $\rho_{11} = 0.9$ and $\rho_{22} = 0.04$ for the other).

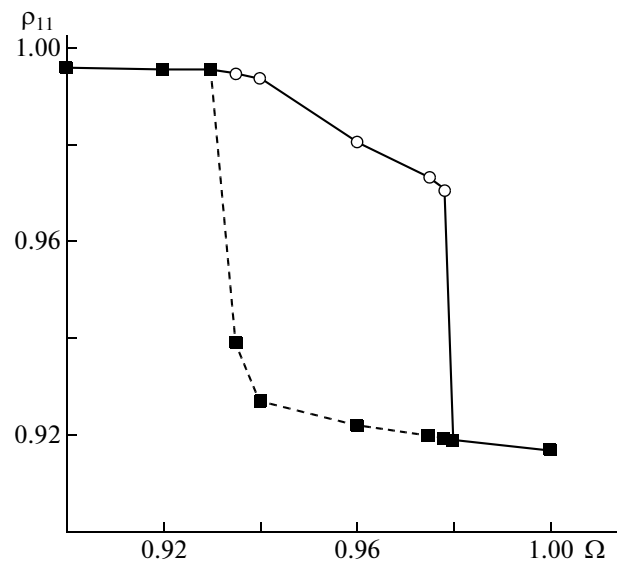

**Fig. 4.** Spatial bistability region for linear *J* aggregate.

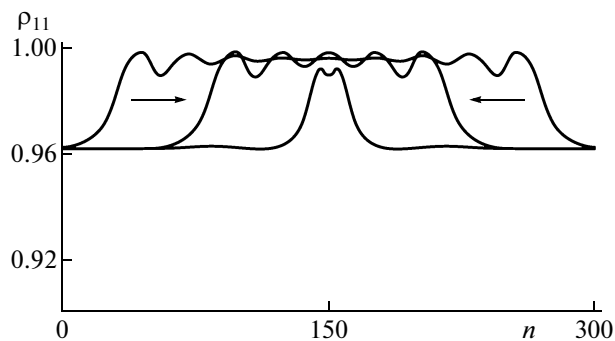

**Fig. 5.** Dynamics of lower level population in the regime of two counter-propagating switching waves (instants $t = 0$, 120, and 440) for ring-shaped *J* aggregate; $N = 300$ and $\Omega = 0.95$.

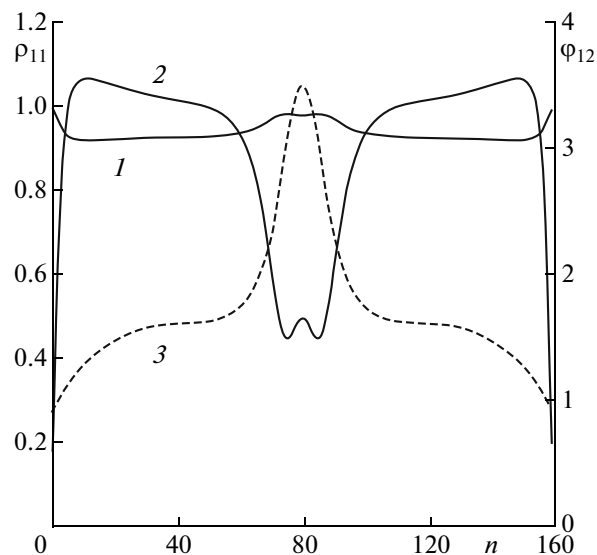

**Fig. 6.** Population of the lower (curve *1*) and upper (curve *2*, magnified by a digit of 40) levels (left scale) and phase $\rho_{12}$ (dashed curve *3*, right scale) for soliton state. $\Omega = 0.95$, $N = 160$.

### 3.3. Switching Waves

In a ring-shaped *J* aggregate with the classical bistability (bistability of uniform states), it is possible to create an initial condition responsible for the occurrence of one stable state at one part of the aggregate and of another state at the other part of the aggregate. Further evolution leads to the formation of switching waves, an example of which is shown in Fig. 5. The figure demonstrates two counter-propagating switching waves that do not interact with each other, while the distance between their fronts exceeds the front widths. It is important that the initially wide nonuniformity gradually transforms to the narrow nonuniformity, which corresponds to solitons considered in the next subsection. A change in the parameters leads to a change in the properties of the switching waves and in the character of their interaction.

### 3.4. Discrete Dissipative Molecular Solitons

This type of soliton corresponds to the profiles of molecular characteristics localized, due to the balance between the energy inflow and outflow, on scales considerably smaller than the scale of the entire aggregate. In a fairly long aggregate, it is possible to excite a large number of solitons with almost arbitrary positions (determined by the excitation conditions). As is known [11], the bistability is a favorable factor for the existence of dissipative solitons. This is also true for molecular aggregates with the ring and chain geometries.

**Ring-shaped *J* aggregates.** The region of existence of discrete dissipative solitons is smaller than the bistability region; solitons exist in the region of about $0.94 \leq \Omega \leq 0.985$ (for $N = 200$). At smaller $\Omega$, the only one stable state is the state with a smaller population of excited molecular states, and state with a larger population is the only one stable state at larger $\Omega$. Within the indicated region, in addition to the two uniform states, there also exist soliton states. Figure 6 shows the distribution of populations of the ground (curve *1*) and excited (curve *2*) levels of molecules along the aggregate, as well as the distribution of the phase of the off-diagonal density matrix element $\rho_{12}$, which determines the molecular dipole moment (curve *3*). This stable spatially-nonuniform state is indeed a soliton, since, by varying the initial distribution, we can form it in different parts of the ring or create several such solitons in the case of a sufficiently long ring. These solitons were not observed in short aggregates (at $N < 90$, which is obviously determined by a decrease in the molecular interaction coefficients $\gamma_{lk}$ and $\Delta_{lk}$ with increasing distance between molecules).

**Linear aggregates.** Similar to the case of ring-shaped aggregates, solitons also exist in linear aggregates (molecular chains) almost in the same region of $\Omega$. One- and many-soliton states can be implemented with different positions on the chain if the chain is sufficiently long (Fig. 7).

### 3.5. Inhomogeneous Broadening

Relation (5) takes into account the arbitrary spread of frequencies of aggregate molecules due to their interaction with the environment, these frequencies being time-independent at low temperatures. This circumstance leads to the violation of the phase coherence of molecules interacting with radiation. In this case, the *J* aggregate is efficiently decomposed into finite-length clusters whose molecules are coherent, while the clusters are incoherent with each other. Generally speaking, the coherence length for a molecular aggregate can be defined in different ways. Since we are primarily interested in the interaction of the aggregate with the radiation, it is convenient to define coherent molecules as the molecules for which the phases of the complex off-diagonal matrix element $\rho_{12}$ determining the dipole moment differ from the field phase by a value smaller than $\pi/2$.

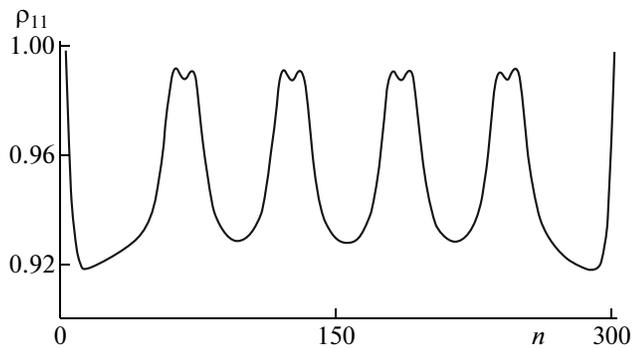

**Fig. 7.** Four-soliton state in a chain of 300 molecules. Rabi frequency is $\Omega = 0.91$.

Let us determine the average cluster size (coherence length) as follows. For each random frequency spread realization (with the realization number $i$), we find a stable histogram, i.e., the dependence of the number of clusters with the size $k$ on $k$, $C_i(k)$. Averaging $C_i(k)$ over all realizations (12 realizations in this calculation), we obtain the function $\text{Clas}(k)$. The average cluster size (in units of the intermolecular distance $a$) or the coherence length $L_c$ can be found by the formula

$$L_c = \sum_1^N k \text{Clas}(k) / \sum_1^N \text{Clas}(k). \qquad (9)$$

We studied the dependence of the average cluster size $L_c$ on the detuning spread dispersion $\sigma$ for different initially (at $\sigma = 0$) uniform states of the $J$ aggregate. We varied the Rabi frequency $\Omega$ from 0.1 to 2 and the frequency detuning dispersion $\sigma$ from 0 to 0.2 (in units of $|\Delta_L|$). For each of these cases, we took 12 realizations of the random addition to the frequency detuning $\Delta_k$ and calculated the coherent cluster size averaged over all the realizations.

As can be seen from Fig. 8, the coherence length (cluster size) decreases with increasing dispersion and depends on the Rabi frequency, which relates to the synchronization of oscillations of molecular dipoles by laser radiation. Note that a faster decrease in the coherence length with increasing $\sigma$ is observed in the region of the nominal bistability (in the absence of frequency dispersion, this region is represented by the curves for the Rabi frequencies $\Omega = 0.96$ and $\Omega = 1$). This can be qualitatively explained by the decomposition of aggregates into clusters corresponding to different branches of the bistable response.

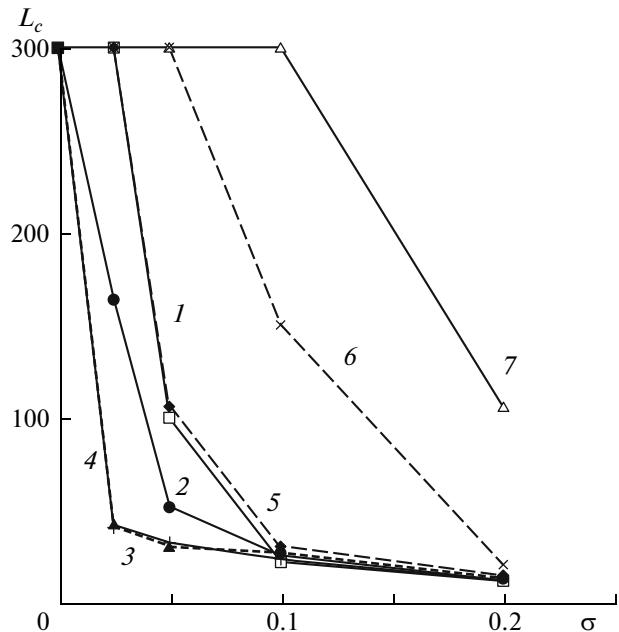

**Fig. 8.** Dependence of averaged (over 12 realizations) coherence length $L_c$ on frequency spread dispersion $\sigma$ for the Rabi frequencies $\Omega =$ (*1*) 0.1, (*2*) 0.9, (*3*) 0.96, (dashed line *4*) 1, (dashed line *5*) 1.1, (*6*) 1.3, and (*7*) 2.

## 4. CONCLUSIONS

Thus, the model of interaction of laser radiation with single linear $J$ aggregates can be corrected in some aspects. First, the methods of modern quantum chemistry allow one to consistently calculate aggregate characteristics, such as the equilibrium geometry and the energies and intensities of the lowest singlet electronic transitions. These data can be used as parameters of a model describing the interaction of $J$ aggregates with radiation in the approximation of the one-particle density matrix. Second, within this model, it is possible to naturally take into account the relaxation processes, the annihilation of excitations at neighboring molecules, and the inhomogeneous broadening. Since the first-principle calculation of the annihilation rate constant is difficult to perform at present, this characteristic can be found from estimates or experimental data.

The calculations performed in this study for molecular chains in the absence of inhomogeneous broadening demonstrate the existence of spatial bistability (when one of two variants of the electronic excitation distribution over the chain is implemented depending on the prehistory), molecular switching waves (with moving fronts of switching between two states), and stationary dissipative solitons. The nonuniform (over the molecular chain length) regimes are obtained under the conditions of a comparatively weak excitation (condition (6)), which allows one to use low-intensity laser radiation. At the same time, the laser should have a stable frequency to retain the nonlinear resonance conditions.

The calculations also revealed the important role of inhomogeneous broadening due to the statistical distribution of frequencies of the main molecular electronic transition due to the interaction of molecules with the environment. This factor leads to an efficient clusterization of *J* aggregates, i.e., to decomposition of the aggregate into separate clusters, inside which the molecules interact with radiation with approximately the same phase difference. A dependence of the average length of these clusters (coherence length) on the frequency spread dispersion and radiation intensity is found. These data show the degree to which it is necessary to decrease the spread dispersion (which can be done by changing the technology of preparation of *J* aggregates) to obtain bistability and solitons.

## ACKNOWLEDGMENTS

This study was supported by the Federal Target Program "Investigations and Developments in the Priority Directions of the Development of Scientific-Technological Complex of Russia 2007–2012" (project no. 2008-04-2.4-15-003), by the Federal Agency for Science and Innovations (State contract no. 02.740.11.0390), by the Russian Foundation for Basic Research (project no. 09-02-92481-MNKS_a, and by the Ministry of Education and Science of the Russian Federation (project no. RNP2.1.1/4694).

*Translated by M. Basieva*